\documentstyle[prb,aps,twocolumn,psfig]{revtex}

\begin{document}
\twocolumn[\hsize\textwidth\columnwidth\hsize\csname
@twocolumnfalse\endcsname

\title {Angle-resolved photoemission study of untwinned
PrBa$_2$Cu$_3$O$_7$: undoped CuO$_2$ plane and doped CuO$_3$ chain} 
 \author{T.~Mizokawa$^1$, C.~Kim$^2$, Z.-X.~Shen$^2$,
A.~Ino$^1$, A.~Fujimori$^1$, M.~Goto$^3$, H.~Eisaki$^3$, S.~Uchida$^3$,
M.~Tagami$^4$, K.~Yoshida$^4$, A.~I.~Rykov$^4$,
Y.~Siohara$^4$, K.~Tomimoto$^4$ and S.~Tajima$^4$} 
 \address{$^1$ Department of Complexity Science and Engineering and
 Department of Physics, University of Tokyo, Bunkyo-ku, 
 Tokyo 113-0033, Japan}
 \address{$^2$ Department of Applied Physics and Stanford Synchrotron 
 Radiation Laboratory, Stanford University, Stanford, CA94305 , U.S.A.}
 \address{$^3$ Department of Superconductivity, University of Tokyo, 
 Bunkyo-ku, Tokyo 113-0033, Japan}
 \address{$^4$ Superconductivity Research Laboratory,
International Superconductivity Technology Center, Koto-ku, Tokyo
135-0062, Japan}
 \date{\today}
\maketitle

\begin{abstract}
We have performed an angle-resolved photoemission study
on untwinned PrBa$_2$Cu$_3$O$_7$, which has low resistivity but
does not show superconductivity. 
We have observed a dispersive feature with a band maximum 
around ($\pi$/2,$\pi$/2), indicating that
this band is derived from the undoped CuO$_2$ plane.
We have observed another dispersive band 
exhibiting one-dimensional character,
which we attribute to signals from the doped CuO$_3$ chain.
The overall band dispersion of the one-dimensional band
agrees with the prediction of $t-J$ model calculation
with parameters relevant to cuprates
except that the intensity near the Fermi level is
considerably suppressed in the experiment.

\end{abstract}


]

\section{Introduction}

In the study of YBa$_2$Cu$_3$O$_7$ (YBCO) and its family
cuprates, the physical properties of the CuO$_3$ chain itself
have been a subject of interest as well as those of the CuO$_2$ plane.
In particular, evidence for charge instability
has been observed in the hole-doped CuO$_3$ chain of YBCO
and its relation with superconductivity has been discussed. \cite{CDW}
Among the YBCO family compounds, PrBa$_2$Cu$_3$O$_7$ (PBCO)
is unique in that it does not show
superconductivity while the other rare-earth
substituted YBCO compounds are superconducting. \cite{Pr123}
Optical studies \cite{Takenaka} have revealed that the CuO$_2$ plane
is not doped with holes and consequently that the superconductivity
is suppressed in PBCO. In order to explain the hole depletion
in the CuO$_2$ plane, several models have been proposed.
Among them, the model proposed by Fehrenbacher and Rice (FR),
\cite{FR} in which the Pr 4$f$-O 2$p_{\pi}$ states trap holes
from the Cu 3$d$-O 2$p_{\sigma}$ band, has been most successful.
It was also argued that
hole transfer between the CuO$_2$ plane and the CuO$_3$ chain
may play an important role. \cite{Khomskii}
Experimental studies of the band structure of
the CuO$_2$ plane, the CuO$_3$ chain and the Pr 4$f$ state
should further reveal the 
difference between the superconducting YBCO and non-superconducting
PBCO. \cite{ETL}
Another important and potentially even more interesting point is
that, if the CuO$_2$ plane is undoped and the CuO$_3$
chain is doped as suggested by the various experimental
and theoretical studies, PBCO may give us
an opportunity to study a hole doped CuO$_3$ chain
with good one-dimensionality compared to YBCO.
Recently,
an angle-resolved photoemission spectroscopy (ARPES)
study of SrCuO$_2$ by Kim {\it et al.} \cite{Kim}
has shown the spinon and holon bands
of the undoped CuO$_3$ chain,
which is a manifestation of spin-charge separation
in the one-dimensional system,
and has attracted much interest.
However, no ARPES study of a hole-doped CuO$_3$ chain
has been made so far.
Although YBCO
has been studied by ARPES, \cite{YBCO}
the band dispersion from the CuO$_3$ chain is
far from being one dimensional because of
the strong interaction between the chain and the plane.

In this paper, we report an ARPES study of untwinned
PBCO samples which have relatively low resistivity but is not superconducting.
As we show below, we observe at least two kinds
of dispersive features: one from
the undoped antiferromagnetic CuO$_2$ plane and
the other which is highly one-dimensional band
from the doped CuO$_3$ chain.

\section{Experimental}

\begin{figure}
\psfig{figure=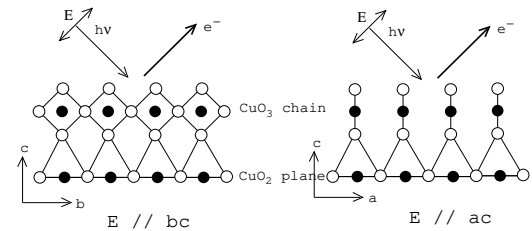,width=7cm}
\caption{
$E \parallel bc$ and $E \parallel ac$
arrangements in which ARPES measurements
were performed. The chains are along the $b$-axis.
}
\label{arrange}
\end{figure}

Single crystals of PBCO 
were grown in a MgO
crucible by a pulling technique. The composition was
determined to be Pr$_{1.02}$Ba$_{1.98}$Cu$_{2.92}$Mg$_{0.08}$O$_7$
by inductively coupled plasma analysis. \cite{ICP}
Because a small amount of Mg-impurities
originating from the crucible are
substituted preferentially for the plane Cu sites, \cite{Mg}
it is expected
that the CuO$_3$ chain is not affected by these impurities. 
Rectangular shaped samples were cut out from the as-grown
crystal and annealed at 500 $^\circ$C in oxygen atmosphere under
uniaxial pressure.
The resistivity parallel to the chain direction
increases with cooling from 8 m$\Omega$cm at 300 K
up to 10 $\Omega$cm at 10K
and that perpendicular to the chain direction
increases from 400 m$\Omega$cm at 300 K
to 70 $\Omega$cm at 10K.
The ARPES measurements were performed using the Vacuum Science
Workshop chamber attached to the undulator beamline 5-3
of Stanford Synchrotron Radiation Laboratory (SSRL).
Incident photons were linearly polarized and had 
energy of 29 eV. The total energy resolution including the 
monochromator and the analyzer was approximately 40 meV. The 
angular resolution was $\pm 1$ degree, which gives the momentum
resolution of $\pm$0.05$\pi$ at $h\nu$ = 29 eV.
[In the text, momenta along the $a$- and $b$-axes are given
in units of $1/a$ and $1/b$, respectively. Here,
$a$ (= 3.87 $\stackrel{\circ}{\rm A}$) and 
$b$ (= 3.93 $\stackrel{\circ}{\rm A}$) are the lattice constants 
of PBCO perpendicular
and parallel to the chain direction, respectively.]
The chamber pressure during the measurements
was less than $5\times 10^{-11}$ Torr.
The samples were cooled to 10 K and cleaved {\it in situ}.
The cleaved surface was the $ab$ plane.
Orientation of the $a$- and $b$-axes was done
by Laue diffraction before and after the measurement. 
The cleanliness of the surfaces was checked by the absence of
a hump at energy $\sim -9.5$ eV. \cite{YBCO}
All the spectra presented here were taken within 
30 hours of cleaving. We cleaved the samples three times
and checked the reproducibility.
As shown in Fig.~\ref{arrange}, 
ARPES data were taken in two arrangements:
an $E \parallel ac$ arrangement
in which the photon polarization $E$ is in the $ac$ plane
and is perpendicular to the chain
and an $E \parallel bc$ arrangement
in which $E$ is in the $bc$ plane
and has a component parallel to the chain.
The position of the Fermi level ($E_{\rm F}$)
was calibrated with gold spectra for every measurement.

\section{Results and discussion}

\begin{figure}
\psfig{figure=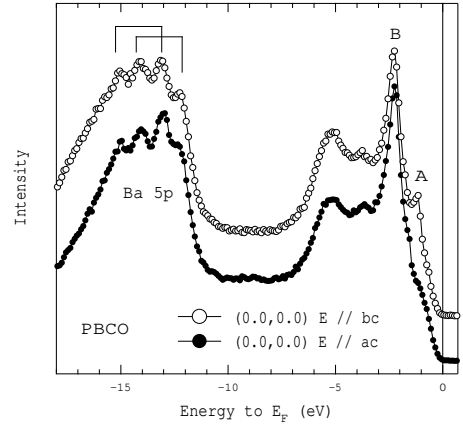,width=7cm}
\caption{
Valence-band spectra taken at (0,0)
with $E \parallel bc$ (open circles)
and that with $E \parallel ac$ (closed circles).
}
\label{val}
\end{figure}

The entire valence band data measured at ($k_a$,$k_b$) = (0,0)
are shown in Fig.~\ref{val}. Here, $k_a$ and $k_b$ are
momenta along the $a$- and $b$-axes, respectively.
The Ba 5$p$ core levels are split into surface and bulk components
as observed in YBCO, \cite{YBCO} indicating that the cleaved surface
is as good as that of YBCO. The peak labeled as A is
intense for $E \parallel bc$, 
while it loses its weight for $E \parallel ac$.
This strong polarization dependence indicates that peak A is
derived from the CuO$_3$ chain and that the CuO$_3$ chain
is well aligned at the surface.
Peak A can be attributed to the non-bonding O 2$p$ states
of the CuO$_3$ chain. \cite{Kim}
On the other hand, the intensity of peak B is insensitive to 
the photon polarization, indicating that
it is derived from the CuO$_2$ plane. Peak B has
energy of $\sim$ -2.3 eV and is similar to
that found in Sr$_2$CuO$_2$Cl$_2$. \cite{Sr2CuO2Cl2}
As pointed out by Pothuizen {\it et al.}, \cite{Sr2CuO2Cl2}
peak B corresponds to the non-bonding O 2$p$ states
of the CuO$_2$ plane.

\begin{figure}
\psfig{figure=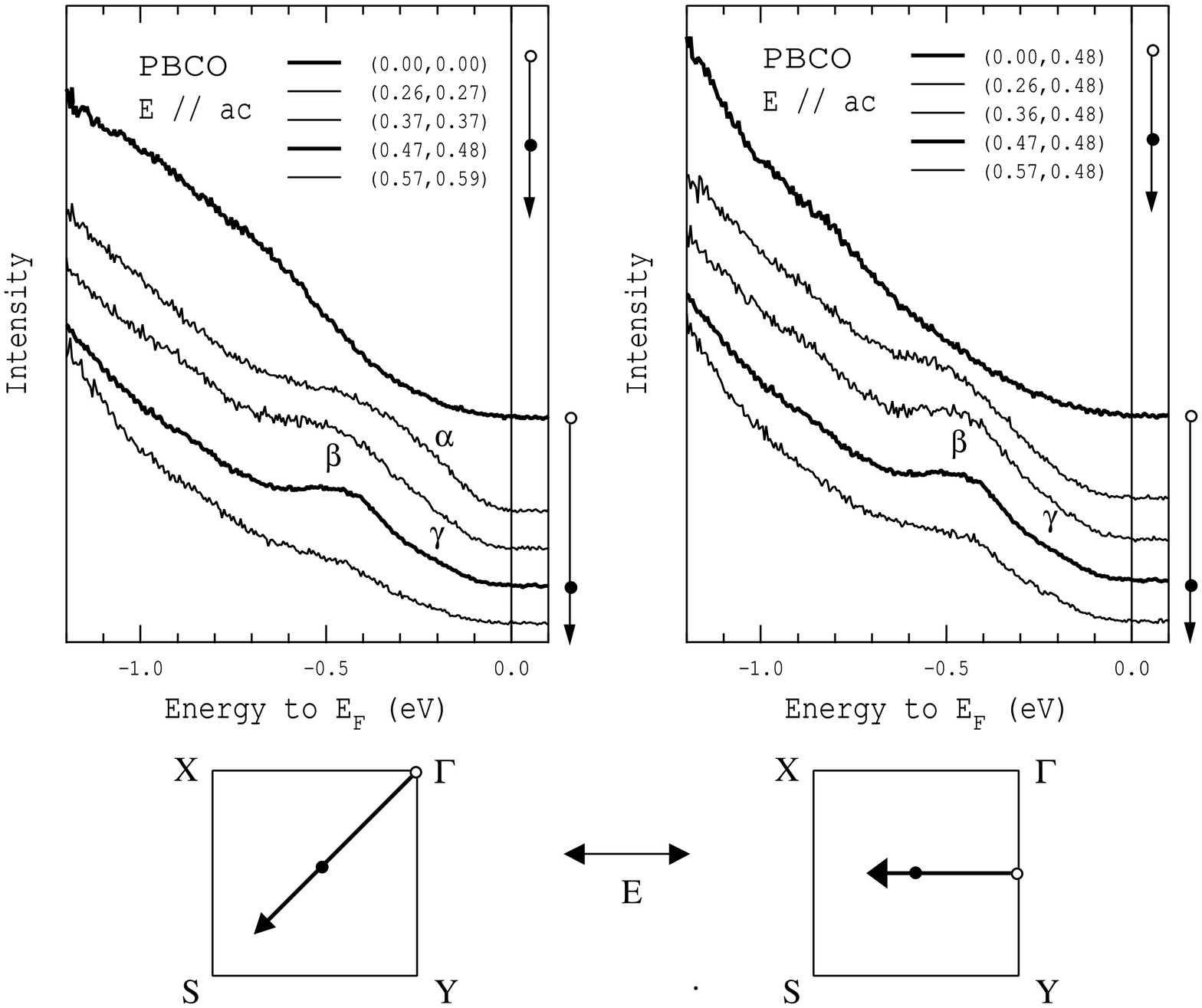,width=7cm}
\caption{ 
Left panel: ARPES spectra along the (0,0)$\rightarrow$(1,1)
direction for $E \parallel ac$.
Right panel: ARPES spectra along the (0,1/2)$\rightarrow$(1/2,1/2)
direction for $E \parallel ac$.
Lower panel shows some measured points (open and closed circles)
and directions in the momentum space and
the in-plane component of the photon polarization (arrows).
$\Gamma$Y is the chain direction.
}
\label{pp1}
\end{figure}

\begin{figure}
\psfig{figure=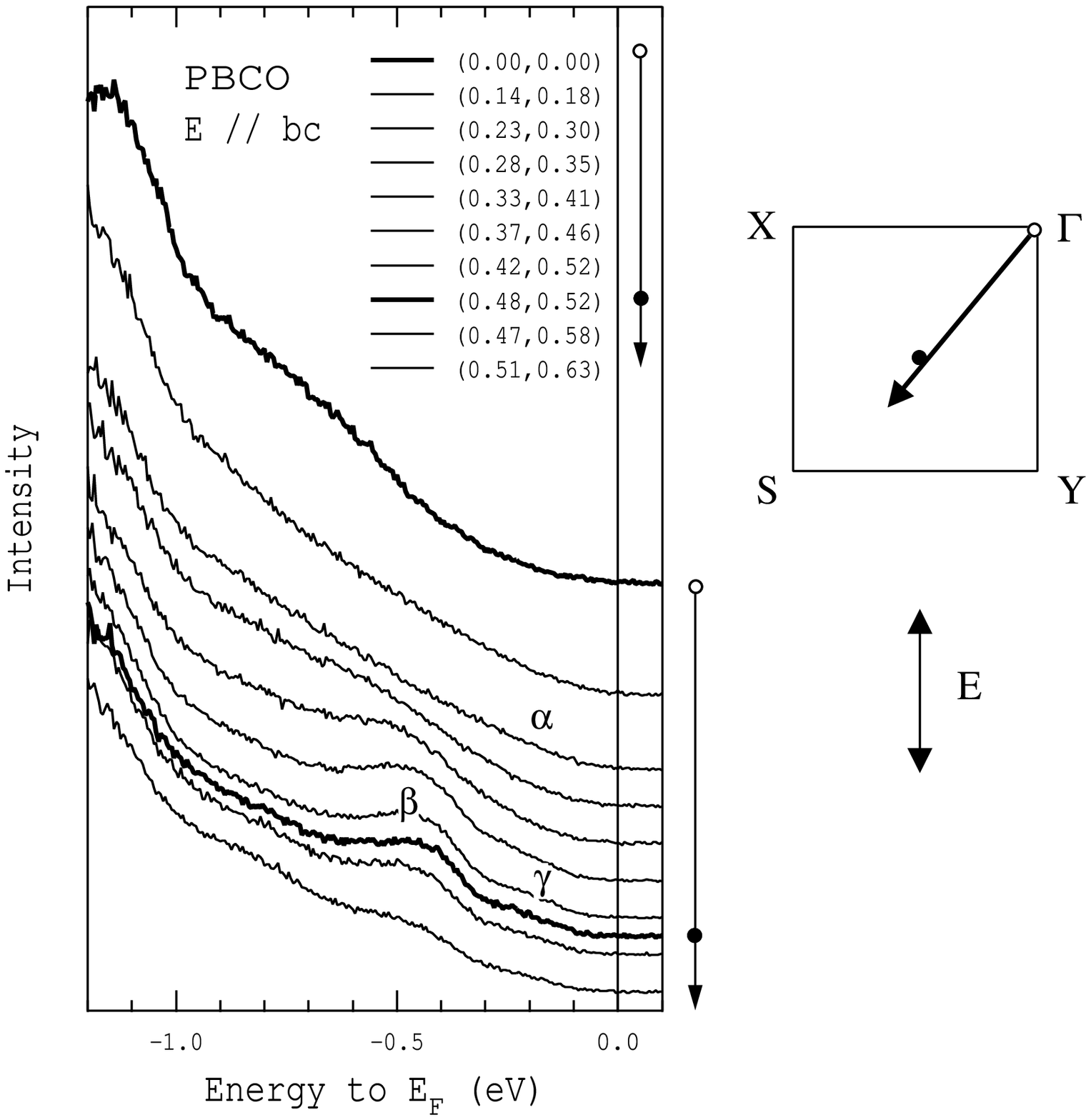,width=7cm}
\caption{ 
ARPES spectra approximately along the (0,0)$\rightarrow$(1,1)
direction for $E \parallel bc$.
The arrows indicate the measured directions.
The momenta along the $a$- and $b$-axes are given
in units of $\pi/a$ and $\pi/b$
Right panel shows some measured points (open and closed circles)
and directions in the momentum space and
the in-plane component of the photon polarization (arrows).
$\Gamma$Y is the chain direction.
}
\label{pp2}
\end{figure}

Figure~\ref{pp1} shows ARPES spectra
along the (0,0)$\rightarrow$($\pi$,$\pi$) direction
and those along the (0,$\pi$/2)$\rightarrow$($\pi$,$\pi$/2) direction
taken with $E \parallel ac$. The uppermost and second uppermost 
spectra in the left panel
were taken at (0,0) and ($\pi$/4,$\pi$/4), respectively.
The intensity from -0.2 to -0.4 eV is enhanced 
at ($\pi$/4,$\pi$/4) compared to that at (0,0). 
Let us denote this feature as $\alpha$ and discuss it
in following paragraphs.
The spectra taken at ($\pi$/2,$\pi$/2) are shown by the thicker
solid curves and marked by the closed circles
in the two panels of Fig.~\ref{pp1}. 
Figure \ref{pp2} shows ARPES spectra
nearly along the (0,0)$\rightarrow$($\pi$,$\pi$) direction
for $E \parallel bc$. In Fig.~\ref{pp2}, the spectrum
taken at ($\pi$/2,$\pi$/2) is shown by the thicker
solid curve and marked by the closed circle.
In these spectra taken around ($\pi$/2,$\pi$/2), 
a dispersive feature with a band maximum at $\sim$ -0.4 eV 
is clearly seen and is labeled as $\beta$.
This structure $\beta$ is very similar to that found
at ($\pi$/2,$\pi$/2) in Sr$_2$CuO$_2$Cl$_2$
and can be interpreted as the Zhang-Rice (ZR)
singlet state of the undoped antiferromagnetic CuO$_2$ plane.
\cite{Sr2CuO2Cl2}
The energy difference between the ZR state at ($\pi$/2,$\pi$/2)
(structure $\beta$) and the non-bonding O 2$p$ state at (0,0) (peak B)
is $\sim$ 2 eV in PBCO, in agreement with the observation
in Sr$_2$CuO$_2$Cl$_2$. \cite{Sr2CuO2Cl2} 
In addition to the ZR state, a weak structure $\gamma$ at -0.2 eV
was observed around ($\pi$/2,$\pi$/2).

By comparing Figs.~\ref{pp1} and \ref{pp2}, one can notice that
the intensity of structure $\alpha$
at ($\pi$/4,$\pi$/4) is weak for $E \parallel bc$
compared to that for $E \parallel ac$.
In addition, while the relative intensity of
structure $\alpha$ to structure $\beta$
has strong polarization dependence,
that of structure $\gamma$ to structure $\beta$
has only small polarization dependence.
This suggests that these two
structures $\alpha$ and $\gamma$ have different origins.
As discussed in the next paragraph,
structure $\alpha$ at ($\pi$/4,$\pi$/4),
which shows strong polarization dependence,
is part of a one-dimensional band from the CuO$_3$ chain.
On the other hand, since structure $\gamma$
shows weak polarization dependence and probably has
two-dimensional character,
it is tempting to interpret structure $\gamma$
as a Pr 4$f$-O 2$p_{\pi}$ hybridized state or a so-called FR state.
If this is the FR state, the present spectra are consistent with the FR
scenario because the FR state is closer to $E_{\rm F}$ than
the ZR state of the CuO$_2$ plane. \cite{XAS} 
The situation that the FR state is partially occupied 
and is observable in photoemission spectroscopy
is consistent with the optical result, which has shown that
the formal valence of the Pr ion is +3.5 and the FR state is
occupied by 0.5 electrons on the average 
if it exists. \cite{Takenaka} This argument on structure $\gamma$
should be confirmed by using Pr 4$d$-4$f$
resonant photoemission in the future.

\begin{figure}
\psfig{figure=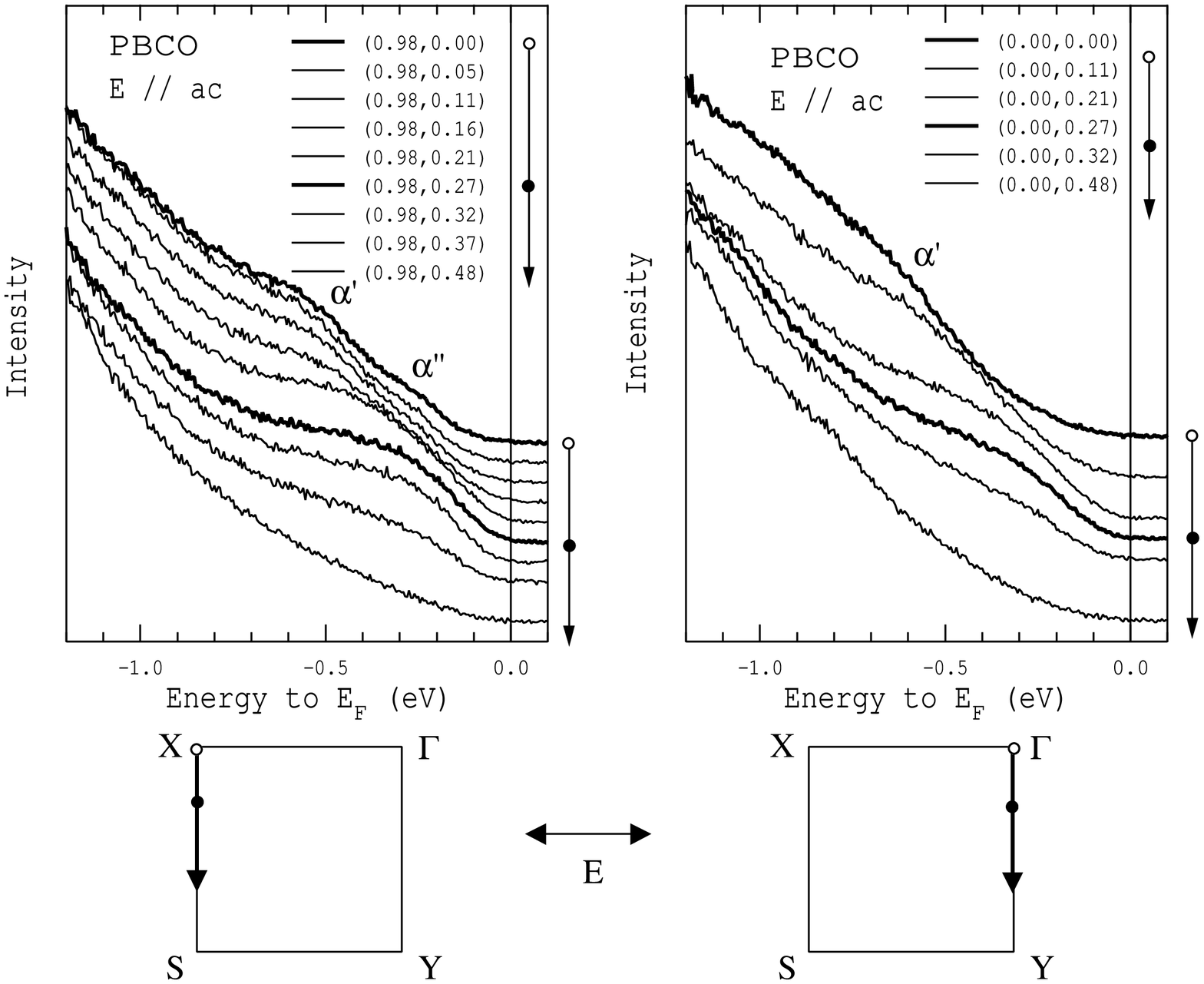,width=8cm}
\caption{
Left panel: ARPES spectra along the (0,0)$\rightarrow$(0,1/2)
direction or the chain direction for $E \parallel ac$. 
Right panel: ARPES spectra along the (1,0)$\rightarrow$(1,1/2)
direction or the chain direction for $E \parallel ac$.
Lower panel shows some measured points (open and closed circles)
and directions in the momentum space and
the in-plane component of the photon polarization (arrows).
$\Gamma$Y is the chain direction.
}
\label{chain1}
\end{figure}

\begin{figure}
\psfig{figure=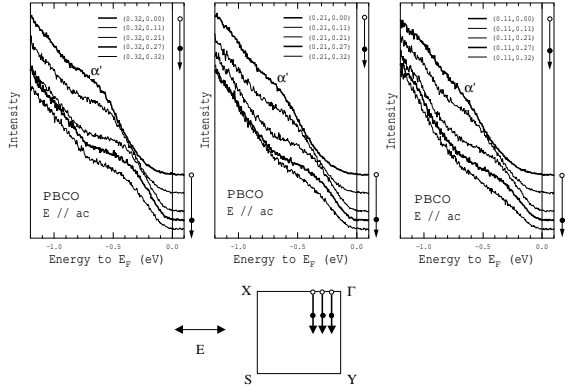,width=8cm}
\caption{
ARPES spectra along the chain direction
taken in the arrangement of $E \parallel ac$.
$k_a$ is the momentum perpendicular to the chain.
Lower panel shows some measured points (open and closed circles)
and directions in the momentum space and
the in-plane component of the photon polarization (arrows).
$\Gamma$Y is the chain direction.
}
\label{chain2}
\end{figure}

\begin{figure}
\psfig{figure=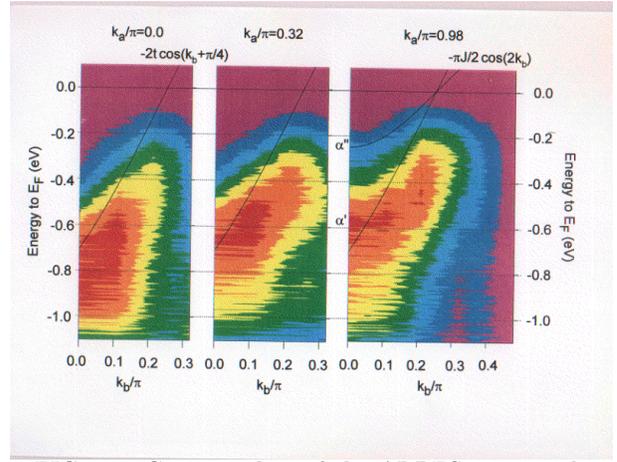,width=8cm,clip=,angle=90}
\caption{
Contour plots of the ARPES spectra 
along the chain direction for $E \parallel ac$. 
Intensity increases in going from blue to yellow to red regions.
}
\label{chain3}
\end{figure}

In Figs.~\ref{chain1} and \ref{chain2}, we have plotted ARPES data
along the chain direction taken with $E \parallel ac$.
In the spectra taken at ($k_a$,0) 
($k_a$ is 0, $\pi$/10, $\pi$/5, 3$\pi$/10, or $\pi$), 
which are shown at the uppermost position in each panel, 
the broad feature labeled as $\alpha'$ is located 
at -0.7 $\pm$ 0.2 eV. 
This feature moves toward $E_{\rm F}$ as $k_b$ increases
and reaches a band maximum with energy of $\sim$ -0.2 eV 
around ($k_a$,$\pi$/4).
The spectra taken at ($k_a$,$\pi$/4) are shown
by the thicker solid curves and marked by the closed circles
in each panel of Figs.~\ref{chain1} and \ref{chain2}.
This feature loses its weight for $k_b > \pi/4$ 
and the spectrum at ($k_a$,$\pi$/2) is almost featureless.
This behavior does not depend on $k_a$, i.e., 
the momentum perpendicular to the chain,
meaning that this dispersive feature is highly one dimensional.
In order to demonstrate the dispersion more clearly, 
the contour plots of the difference spectra, which are 
obtained by subtracting the featureless spectrum at ($\pi$,$\pi$/2)
from the spectra at each $k$-point, are shown 
for $k_a$ = 0, 3$\pi$/10, and $\pi$ in Fig.~\ref{chain3}. 
Indeed, the overall dispersion of structure $\alpha'$ 
does not depend on $k_a$ so much.
Therefore,
we can conclude that the band is derived from the CuO$_3$ chain.
Structure $\alpha$ at ($\pi$/4,$\pi$/4)
[see Fig.~\ref{pp1}] is also part of this one-dimensional band.
The fact that the one-dimensional band reaches a band maximum
at $k_b$ $\sim$ $\pi$/4 and disappears for $k_b > \pi/4$
indicates that the filling of 
the Cu 3$d_{x^{2}-y^{2}}$-O 2$p_\sigma$ band in the CuO$_3$ chain
is close to 1/4, namely, the formal valence of Cu in the chain
is $\sim$ +2.5.
This is consistent with the optical study. \cite{Takenaka}

\begin{figure}
\psfig{figure=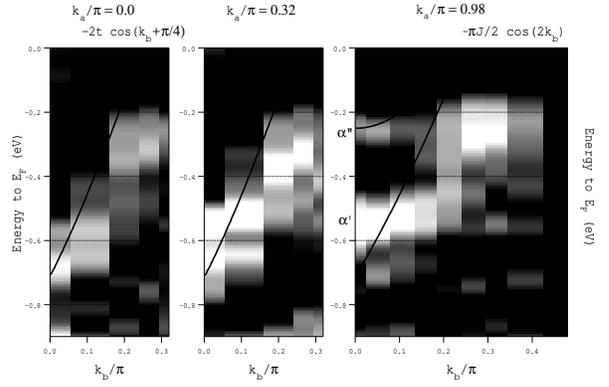,width=8cm}
\caption{
Second derivatives of the ARPES spectra
along the chain direction for $E \parallel ac$.
In the right panel, two dispersive features are visible
as two bright belts.
The solid curves are model holon and spinon dispersions
given by -2$t$cos($k_b$+$\pi$/4) and -$J$cos(2$k_b$)
with $J$ = 0.5 eV and $t$ = 0.16 eV.
}
\label{chain4}
\end{figure}

There are two interesting points in this one-dimensional band.
The first point 
is that the other structure ($\alpha''$), which has higher
energy than structure $\alpha'$, is observed around ($\pi$,0)
[Compare the uppermost spectra shown by the thicker solid curves
in the two panels of Fig.~\ref{chain1}.].
In order to show the dispersion clearly, the density plots
of the second derivatives of the ARPES spectra along the chain
direction are diplayed for $k_a$ = 0, 3$\pi$/10, and $\pi$
in Fig.~\ref{chain4}. The dispersion of $\alpha''$ 
is visible in the spectra for $k_a$ = $\pi$ 
which is shown in the right panel of Fig.~\ref{chain4}.
It is possible to attribute $\alpha'$
and $\alpha''$ to the holon and spinon bands
of the Tomonaga-Luttinger (TL) liquid. \cite{TL}
While the total width of the holon band is predicted
to be $\sim$ 4$t$, that of the spinon band
has an energy scale of $J$. \cite{Maekawa} 
In Figs. \ref{chain3} and \ref{chain4},
model holon and spinon dispersions of -2$t$cos($k_b$+$\pi$/4) and 
-$\pi J/2$cos(2$k_b$) \cite{TL,Maekawa} are shown by solid curves. 
The two curves with $t$ of 0.5 eV and $J$ of 0.16 eV,
which are reasonable values for the cuprates, \cite{Maekawa}
roughly follow the dispersions
in Fig. \ref{chain4}.
The second point is that the spectral weight
near $E_{\rm F}$ is considerably suppressed
in the experimental data in disagreement with the theoretical
prediction on the TL liquid. \cite{TL,Maekawa}
Although the lengths of the CuO$_3$ chain at the surface are 
finite because of the surface termination, \cite{YBCO}
the observation of the nice dispersive behavior 
indicates that the lengths of the CuO$_3$ chains
are long enough to allow us to compare the data 
with the theory for the doped CuO$_3$ chain.
A possible origin of the intensity suppression
near $E_{\rm F}$ is the instability of
the nearly-1/4-filled CuO$_3$ chain leading to
charge density waves (CDW). \cite{Lee}
Actually, charge instability in the hole-doped
CuO$_3$ chain of PBCO has been observed by NMR and NQR
measurements. \cite{Grevin} In addition, 
it has recently been pointed out that
the spectral function of one-dimensional CDW insulators can have 
the holon and spinon dispersions. \cite{Voit}

Here, it should be remarked how we would observe the feature with
$E \parallel ac$. In PBCO,
the CuO$_4$ square plane of the CuO$_3$ chain is perpendicular
to the $ab$ plane, i.e., the sample surface.
Therefore, the photon polarization has a component perpendicular
to the sample surface, namely, parallel to
the CuO$_4$ square plane of the CuO$_3$ chain as shown in Fig.~\ref{arrange}.
It is this additional component of the polarization that gives
a finite transition matrix element
to the ZR state in the CuO$_3$ chain.

\begin{figure}
\psfig{figure=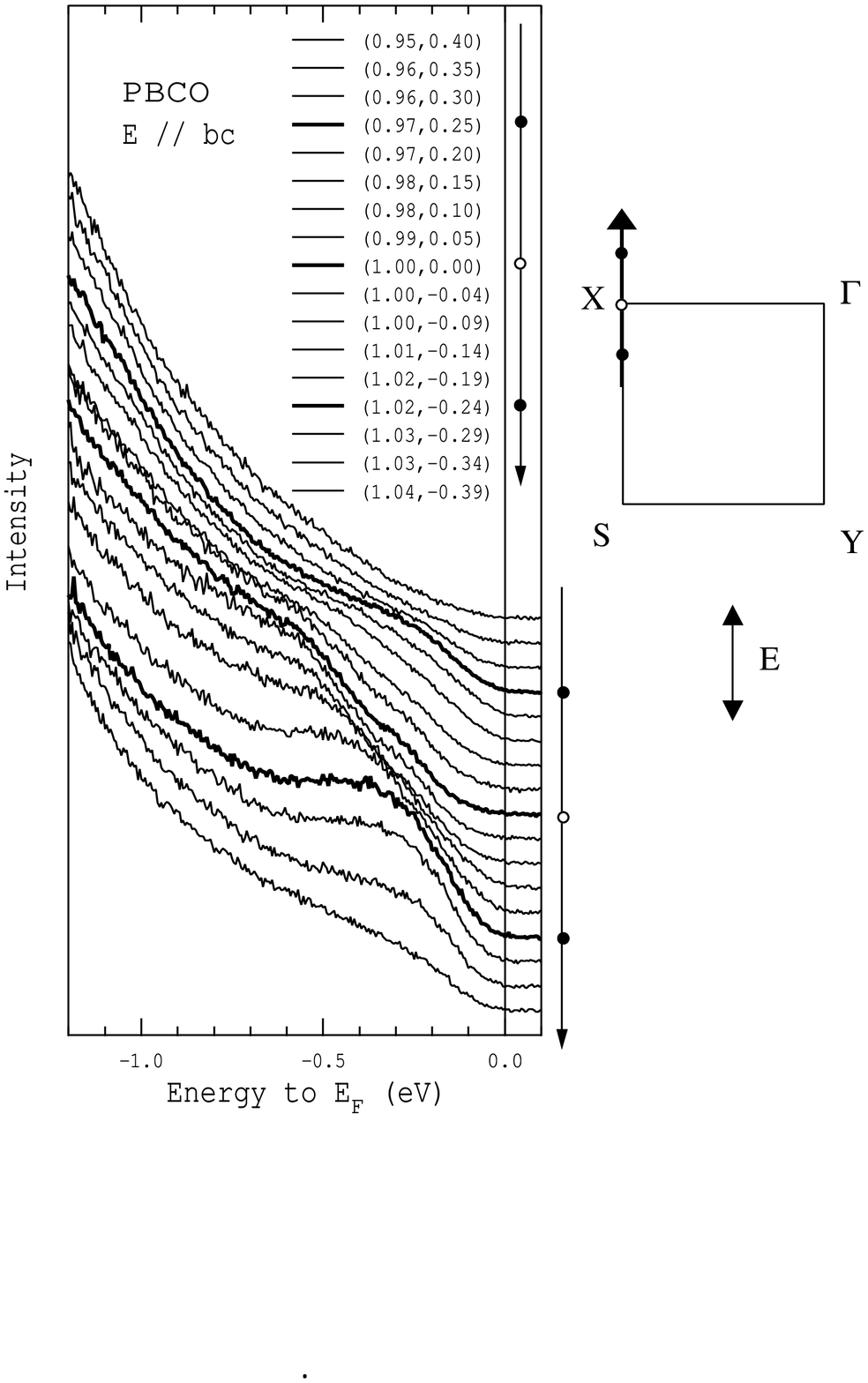,width=7cm}
\caption{
ARPES spectra along the (1,-1/2)$\rightarrow$(1,1/2)
direction or the chain direction for $E \parallel bc$. 
Right panel shows some measured points (open and closed circles)
and directions in the momentum space and
the in-plane component of the photon polarization (arrows).
$\Gamma$Y is the chain direction.
}
\label{chain5}
\end{figure}

\begin{figure}
\psfig{figure=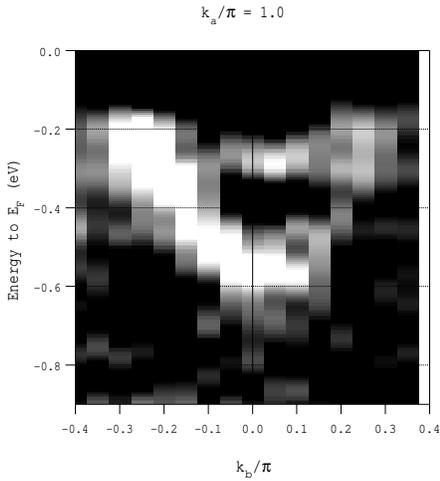,width=7cm}
\caption{
Second derivatives of the ARPES spectra
along the chain direction for $E \parallel bc$.
Two dispersive features are visible as two bright belts.
}
\label{chain6}
\end{figure}

With $E \parallel bc$, the contribution from the chain
is very weak when $k_a$ is small.
As $k_a$ becomes larger, the intensity of the
one-dimensional band for $E \parallel bc$ becomes larger 
and, at $k_a$ = $\pi$, is comparable to that for $E \parallel ac$.
Figure~\ref{chain5} shows ARPES spectra
along the ($\pi$,$\pi$/2)$\rightarrow$($\pi$,-$\pi$/2) direction,
namely, along the chain direction taken with $E \parallel bc$.
The two dispersive features, which can be interpreted
as the holon and spinon dispersions,
are also observed for $E \parallel bc$. 
In order to show the dispersion obtained 
for $E \parallel bc$ clearly, 
the density plot of the second derivatives 
of the ARPES spectra is displayed in Fig.~\ref{chain6}.
The dispersions of the holon and spinon bands 
are almost symmetric with respect to $k_b$ = 0
and reach maxima around $k_b$ = $\pi$/4 and -$\pi$/4.
These dispersions obtained for $E \parallel bc$
are consistent with those obtained for $E \parallel ac$.

In the present ARPES data taken
with $E \parallel ac$ and $E \parallel bc$, 
the relative intensity of the spinon band to the holon band
strongly depends on $k_a$. The spinon band becomes more intense
as $k_a$ becomes larger. The same behavior
was found in the undoped CuO$_3$ chain by Kim {\it et al.} \cite{Kim}
Further experimental and theoretical investigation
is required to reveal this peculiar $k_a$ dependence of the spectral
function.

\section{Conclusion}

In conclusion, we have observed the band dispersions
from the CuO$_2$ plane and 
the hole-doped CuO$_3$ chain of non-superconducting
PBCO. The band dispersion from the CuO$_2$ plane clearly shows that
PBCO has an undoped insulating CuO$_2$ plane. On the other hand,
the one-dimensional dispersive feature from the doped
CuO$_3$ chain consists of two structures which can be interpreted
as holon and spinon bands.
These bands lose intensity beyond $k_b$ $\sim$ $\pi$/4,
indicating that the CuO$_3$ chain is nearly 1/4-filled.
Further experimental and theoretical studies are desirable
to reveal the nature of the one-dimensional band
including its momentum and polarization dependence.

\section*{Acknowledgment}

We would like to thank M. Schabel, I. Terasaki, T. Thoyama,
S. Maekawa, D. D. Sarma, and K. Penc for valuable comments.
We are grateful to P. J. White, A. Y. Matsuura and
the staff of SSRL for technical support.
This work was supported by a Grant-in-Aid for Scientific Research
from the Ministry of Education, Science, Sports and Culture of Japan,
Special Coordination Funds of the Science and Technology Agency of Japan,
the New Energy and Industrial Technology Development
Organization (NEDO), the U.~S.~DOE, Office of Basic Energy 
Science and Division of Material Science. 
SSRL is operated by the U.~S.~DOE, Office of Basic Energy 
Sciences, Division of Chemical Sciences.

\end{document}